\documentstyle[12pt,epsfig,a4wide]{article}
\newcommand {\br}{\mbox{Br}}
\def\etal{{\it et al.}}
\def\tonetwo{\theta_{12}}
\def\ttwothree{\theta_{23}}
\def\tonethree{\theta_{13}}

\def\d13{\delta_{13}}
\def\Vud{{\rm V}_{\rm ud}}
\def\Vus{{\rm V}_{\rm us}}
\def\Vub{{\rm V}_{\rm ub}}
\def\Vcd{{\rm V}_{\rm cd}}
\def\Vcs{{\rm V}_{\rm cs}}
\def\Vcb{{\rm V}_{\rm cb}}
\def\Vtd{{\rm V}_{\rm td}}
\def\Vts{{\rm V}_{\rm ts}}
\def\Vtb{{\rm V}_{\rm tb}}

\def\cVub{{\rm V}^{\star}_{\rm ub}}

\def\cVcs{{\rm V}^{\star}_{\rm cs}}
\def\cVcb{{\rm V}^{\star}_{\rm cb}}

\def\cVts{{\rm V}^{\star}_{\rm ts}}
\def\cVtb{{\rm V}^{\star}_{\rm tb}}
\def\Vij{{\rm V}_{ij}}
\def\dms{\Delta m_{\rm s}}
\def\dmd{\Delta m_{\rm d}}

\def\mbs{{\rm B}^{0}_{\rm s}}
\def\mbd{{\rm B}^{0}_{\rm d}}
\def\mbzero{{\rm B}^{0}}
\def\bsbsbar{{\rm B}^{0}_{\rm s}\bar{\rm B}^{0}_{\rm s}}
\def\bdbdbar{{\rm B}^{0}_{\rm d}\bar{\rm B}^{0}_{\rm d}}
\def\bzbzbar{{\rm B}^{0}\bar{\rm B}^{0}}
\def\Kpi0nunu{{\rm K}_{\rm L}\rightarrow\pi^{\circ} \nu\bar\nu}
\def\Kpinunu{{\rm K}^+\rightarrow\pi^{+} \nu\bar\nu}
\def\lKpinunu{{\rm K}^+\rightarrow\pi^{+} \nu_{\ell}\bar{\nu_{\ell}}}
\def\Ke3{{\rm K}^+\rightarrow\pi^\circ{\rm e}^+\nu_{\rm e}}
\def\GeVcsq{{\rm GeV}/c^2}
\begin{document}
%
%\vspace{0.6in}
 {\title{
         {
          {\bf
            Bayesian estimate of the effect of $\bzbzbar$ mixing 
            measurements on the CKM matrix elements
          }
         }
        }
 }
\author{David E. Jaffe\thanks{e-mail: djaffe@scri.fsu.edu}\ \ and 
        Saul Youssef\thanks{e-mail: youssef@scri.fsu.edu} \\
        Supercomputer Computations Research Institute\\
           Florida State University\\
           Tallahassee, FL 32306-4052}
\maketitle
\abstract{
          A method employing Bayesian statistics 
          is used to incorporate recent
          experimental results on
          $\bdbdbar$ and $\bsbsbar$ mixing into a measurement 
          of the Cabibbo-Kobayashi-Maskawa matrix elements
          with small theoretical uncertainties.
          The neutral B meson mixing results yield a slight improvement
          in the estimate of $|\Vtd|$. Prospects for improving the
          knowledge of the CKM matrix elements with measurements
          of $\bsbsbar$ mixing and the $\Kpinunu$ branching ratio
          are considered.
         }
\vspace{1.5cm}

 The precise determination of the elements 
of the Cabibbo-Kobayashi-Maskawa (CKM) matrix \cite{CKM} 
is one important goal of
current experiments at BNL, CERN, CESR, FNAL, and SLAC and future
experiments such as BaBar, Belle, CLEO-III, HERA-B, KTeV and LHC-B.
Buras \cite{Buras,Buras2} has pointed out that 
measurement of CP
asymmetries in neutral B meson decays and
the $\Kpi0nunu$
branching ratio 
can determine the CKM matrix with
almost no theoretical uncertainties
if $\Vus$ is known.
As yet, no measurements of these quantities exist.
However,
current experiments do provide information on 
two other useful measurements 
which are estimated to have ${\cal O}(10\%)$
theoretical uncertainties:
the fractional $\bzbzbar$ mass difference
%the ratio of the differences of the $\bzbzbar$ mass eigenstates 
%$\bzbzbar$ mixing 
and the $\Kpinunu$ branching ratio.

\section{{{\bf $\bzbzbar$ mixing }}}

\par In the Standard Model, mixing of neutral B mesons occurs via box
diagrams dominated by internal top quark loops
which allow the ratio of the CKM elements 
$\Vts/\Vtd$ to be determined from
$\bdbdbar$ and $\bsbsbar$ mixing measurements \cite{Buras,Ali}:

\begin{equation}
\frac{\dms}{\dmd}
=
\xi^2 \frac{m_{\rm s}}{m_{\rm d}}
\left|\frac{\Vts}{\Vtd}\right|^2
\label{theory}
\end{equation}

\noindent where $\dmd$ ($\dms$) is the mass difference between
the $\mbd$ ($\mbs$) mass eigenstates,
$m_{\rm s}$ ($m_{\rm d}$) is the $\mbd$ ($\mbs$) mass 
and $\xi = 1.16 \pm 0.10$ \cite{Ali} is the ratio
of hadronic matrix elements for the $\mbs$ and $\mbd$ mesons and
constitutes the theoretical uncertainty 
due to ${\rm SU}(3)_{\rm flavour}$ breaking effects. 
Currently, $\bdbdbar$ mixing is rather well measured, 
$\dmd = 0.457 \pm 0.019 \ {\rm ps}^{-1}$ \cite{SauLan};
while experimental information from LEP on $\bsbsbar$ mixing
%from the behavior of likelihoods ${\cal L}(\dms)$
indicates that $\dms$ is at least ten times larger 
than $\dmd$ \cite{forty,wisconsin,dsl,delphi,opal}.
Although the LEP experiments are incapable of resolving 
%large values of $\dms$ 
values of $\dms$ greater than $10 \ {\rm ps}^{-1}$
due to their limited event samples and proper time resolution,
we show that it is nonetheless possible to determine the 
additional constraints placed
on the CKM matrix elements from the experimental information
on B meson mixing using Bayes' theorem.
By way of demonstration, we first employ Bayes' theorem to determine 
the allowed range of $\dms$ given equation \ref{theory}
and prior measurements of the
magnitudes of CKM matrix elements \cite{PDG}. 

\par
 For a proposition of interest $q$, prior knowledge $i$ and 
additional information $e$, Bayes' theorem 
\begin{equation}
P(q|i\wedge e) = P(q|i) \ P(e|i\wedge q)/P(e|i)
\end{equation}
allows one to adjust the ``prior probability" of $q$,  $P(q|i)$,
given the additional information $e$ \cite{notation}.  
The factor $P(e|i\wedge q)$
is called the ``likelihood'' and often factorizes further since
$e$ is frequently the result of a sequence of statistically 
independent measurements and $P(e|i)$ can often be determined by 
normalization \cite{Jaynes,Jaynes2,Cox}.  In our case, we use Bayes' theorem
to demonstrate how probability densities related to the CKM matrix are
affected by recent $\bzbzbar$ mixing results and how these 
probabilities might further be affected by potential measurements
of the $\Kpinunu$ branching ratio.
\par
  Using the ``standard'' parameterization of the CKM matrix 
as recommended by the Particle Data Group (PDG) \cite{PDG}, 
the CKM matrix is determined by four angles 
$\Theta=(\tonetwo,\ttwothree,\tonethree,\d13)$.  Assuming a 
uniform prior for $\Theta$ within $[0,\pi/2)^3\times [0,2\pi)$,
the new probability density for $\Theta$ is 
\begin{equation}
P(\Theta) = \prod_{i=1}^6 g(v_i,\mu_i,\sigma_i)
\label{Peqn}
\end{equation}
where $g(v,\mu,\sigma)$ is the Gaussian distribution and 
$v_1,v_2,\dots,v_6$ are the six quantities listed in Table \ref{table1}

\begin{table}[h]
\begin{center}
{\tabcolsep 0.5mm
 \begin{tabular}{|c|rcl|}
 \hline
 Matrix element & \multicolumn{3}{c|}{Magnitude} \\
 \hline
 $\Vud$ &\ 0.9736 & $\pm$ & 0.0010\ \\
 \hline
 $\Vus$ &\ 0.2205 & $\pm$ & 0.0018\ \\
 \hline
 $\Vcd$ &\ 0.224 & $\pm$ & 0.016\ \\
 \hline
 $\Vcs$ &\ 1.01 & $\pm$ & 0.18\ \\
 \hline
 ${\Vub}/{\Vcb}$ &\ 0.08 & $\pm$ & 0.02\ \\
 \hline
 $\Vcb$ &\ 0.041 & $\pm$ & 0.003\ \\
 \hline
 \end{tabular}
}
\parbox{15cm}
{
 \caption{
          Experimental determination of the magnitudes of six
          of the CKM matrix elements as compiled by the PDG 
          \protect\cite{PDG}.
         }
 \label{table1}
}
\end{center}
\end{table}

\noindent expressed as functions on $[0,\pi/2)^3\times [0,2\pi)$ 
with independently measured values $\mu_1,\mu_2,\dots,\mu_6$
and corresponding variances $\sigma_1^2,\sigma_2^2,\dots,\sigma_6^2$.
Once we have a probability density $\phi(\Theta)$ then
the density $\psi(x)$ for any real function $f(\Theta)$,
\begin{equation}
\psi(x) = \frac{{\rm d}}{{\rm d} x}
          \int_{\Theta}{\rm d}\Theta\ \phi(\Theta)
          \ \lceil f(\Theta)\leq x\rceil \ ,
\label{pdf}
\end{equation}
can be evaluated numerically \cite{notation,Vegas}.

\begin{figure}[h]
\begin{center}
%\centerline{\epsffile{osc$tmp:3april1.eps}}
%\centerline{\epsfig{file=osc$tmp:8july2.eps,height=12cm}} % was 8cm
\centerline{\epsfig{file=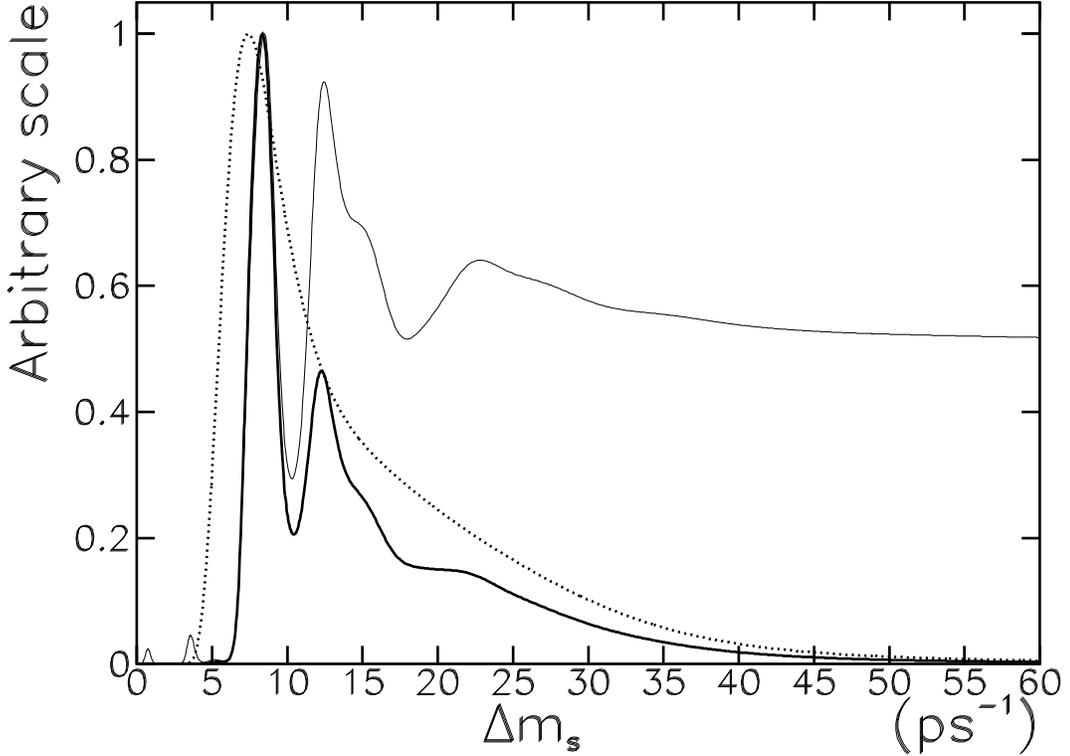,height=12cm}} % was 8cm
% exec CKME2#BAYES_DMS to make this figure
\parbox{15cm}{
\caption{
        % (a) 
        The dotted curve shows the prior probability of 
         \protect$\dms$ determined from the measurements in Table
         \protect\ref{table1}.
         The thin solid curve is \protect${\cal L}(\dms)$
         the likelihood of \protect$\dms$ 
         from the {\sc ALEPH} experiment \protect\cite{dsl} and the thick solid curve
         is the overall probability of \protect$\dms$ determined using
         Bayes' theorem.
         The normalization is arbitrary.
        }
\label{fig1}
}
\end{center}
\end{figure}

  As an example of the method described above, consider the recent {\sc ALEPH}
study of the $\bsbsbar$ mass difference. As shown in Figure \ref{fig1},
the experimental likelihood function ${\cal L}(\dms)$ reflects the inability
of the experiment to distinguish large values of $\dms$ 
and only a lower limit on $\dms$ is reported \cite{dsl}.
However, assuming equation \ref{theory} one 
can use the measurements in 
Table \ref{table1} together with measurements of $\dmd$ \cite{SauLan}, the 
$\mbzero$ meson masses \cite{PDG}
and a theoretical estimate of $\xi$ \cite{Ali} as 
prior information in Bayes' theorem to improve upon the {\sc ALEPH} likelihood.
Assuming that these measurements and theoretical estimates are 
independent and normally 
distributed, the prior probability density of
$\dms$ can be constructed from equation \ref{pdf} given the
constraints in Table \ref{table1}  and is shown in Figure \ref{fig1}.
There is a 95\% probability that $\dms > \ 5.8 \ {\rm ps}^{-1} $ from the
prior probability density alone.
If the prior probability is combined with the {\sc ALEPH} likelihood using
Bayes' theorem, then the lower limit improves to $7.5 \ {\rm ps}^{-1} $ 
as shown in Figure \ref{fig1}.

\begin{figure}[h]
\begin{center}
%\centerline{\epsfig{file=osc$tmp:5april1.eps,height=8cm}}
\centerline{\epsfig{file=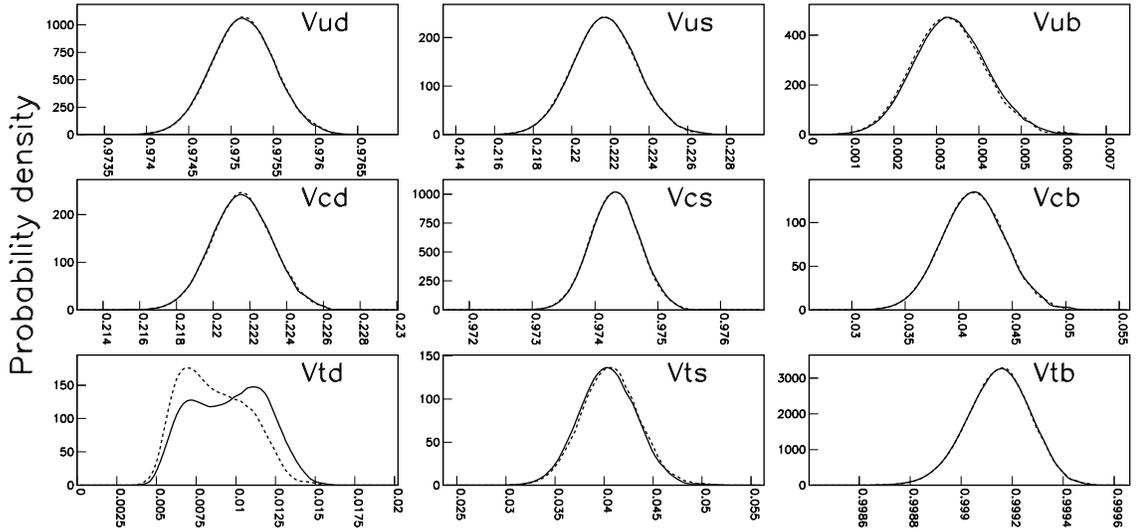,height=8cm}}
\parbox{15cm}{
\caption{
         The probability density functions 
         for the magnitudes of
         the nine CKM matrix elements.
         The solid curves show the probability densities determined
         with the PDG information in Table \protect\ref{table1} alone;
         the dashed curves show the 
         densities with the addition
         of the \protect$\bzbzbar$ mixing results.
        }
\label{fig2}
}
\end{center}
\end{figure}

  The influence of the $\bzbzbar$ mixing measurements on the CKM matrix elements
can be determined if equation \ref{theory} is assumed.  Given the 
{\sc ALEPH} likelihood ${\cal L}(\dms)$ and assuming that $\dmd$, 
$\xi$, $m_{\rm s}$ and $m_{\rm d}$ are independent and Gaussian with
means and variances as quoted above, then equation \ref{pdf} can be
used to determine the probability density for $|\Vts/\Vtd|$ or the 
likelihood $L(\Theta)$ of the {\sc ALEPH} results given
the CKM angles.  Using this likelihood and the PDG prior $P(\Theta)$,
the density $\psi_{ij}$ for the magnitude of CKM matrix element $\Vij$ 
is determined by 
\begin{equation}
\psi_{ij}(v) 
 \propto
 \frac{{\rm d}}{{\rm d} v}
  \int_{\Theta}{\rm d}\Theta \ P(\Theta)
     \ L(\Theta)
     \ \lceil |\Vij(\Theta)|\leq v\rceil
\label{eqnc}
\end{equation}
where the $\Vij(\Theta)$ are the matrix elements expressed
as a functions of the CKM angles.
The resulting densities for the nine CKM elements, both
with and without the $\bzbzbar$ mixing measurements,
are shown in Figure \ref{fig2}.
In the absence of $\bzbzbar$ mixing measurements,
the 90\% confidence
limits on the matrix element magnitudes \cite{central} are

\[
 \left( 
  \begin{array}{ccc}
  \Vud & \Vus & \Vub \\
  \Vcd & \Vcs & \Vcb \\
  \Vtd & \Vts & \Vtb \\
  \end{array}
 \right)
\sim
{\arraycolsep 0.5mm
 \left( 
  \begin{array}{lcllcllcl}
  0.9745 &\mbox{to}& 0.9757&\ 0.219  &\mbox{to}& 0.224 &\ 0.002  &\mbox{to}& 0.005  \\
  0.219  &\mbox{to}& 0.224 &\ 0.9736 &\mbox{to}& 0.9749&\ 0.036  &\mbox{to}& 0.046  \\
  0.006  &\mbox{to}& 0.013 &\ 0.035  &\mbox{to}& 0.045 &\ 0.9989 &\mbox{to}& 0.9993 \\
  \end{array}
 \right)
}\ .
\]

\noindent There is good agreement between these results and 
the PDG \cite{PDG} for all elements except for
$\Vts$, which is due to their 
rounding procedure \cite{Renk},
and $\Vtd$, which is obviously non-Gaussian. The PDG \cite{PDG,Renk} and
others \cite{Schmidtler,Silverman} use the method of least squares
to calculate confidence levels which can lead to inaccurate results
if the predicted distribution of the magnitude 
of a matrix element is non-Gaussian.
The two-lobed structure
of $|\Vtd|$ is due to the phase $\d13$ which is virtually unconstrained 
by the measurements of Table \ref{table1}.

\noindent Only $|\Vtd|$ is significantly changed by the addition 
of the $\bzbzbar$ mixing information. As shown in Figure \ref{fig2},
the 90\% confidence limits change from (0.0057 to 0.0131) to
(0.0054 to 0.0121) as large values of $|\Vtd|$ are disfavoured by
${\cal L}(\dms)$ as shown in Figure \ref{fig1}.

\section{{{\bf The $\Kpinunu$ branching ratio}}}
\par 
%The impact of the experimental and theoretical information
%on the $\Kpinunu$ branching ratio on the CKM matrix can also 
%be determined using Bayes' theorem. 
A Bayesian estimate of the CKM matrix can also be constructed
from $\Kpinunu$ branching ratio measurements together with additional
theoretical assumptions.
The $\lKpinunu$ branching ratio
for a single lepton species $\ell$ is given by \cite{BB}

\begin{equation}
\br(\lKpinunu)
=
\frac{\alpha^2\br(\Ke3)}
     {2\pi^2\sin^4{\theta}_{\rm w}}
\times
\frac{\left|\cVcs\Vcd X^{\ell}_{NL} + \cVts\Vtd X(x_t)\right|^2}
     {\left|\Vus\right|^2} \ ,
\label{Keqn}
\end{equation} 

\noindent where $x_t \equiv m_t^2/m_W^2$.
From Table 1 of reference \cite{BB}, we assume
$X^{\ell}_{NL}  = (11.2 \pm 1.0)\times 10^{-4}$ for $\ell = e,\mu$
and 
$X^{\ell}_{NL}  = (7.6 \pm 1.0)\times 10^{-4}$ for $\ell = \tau$
where the quoted errors are dominated by 
uncertainties in $\Lambda_{\overline{\rm MS}}$ and
the charm quark mass.
We use the approximation 
$X(x_t) = 0.65x_t^{0.575}$ which is accurate
to better than 0.5\% for $150 \le m_t \le 190 \ \GeVcsq$ \cite{Buras2},
and take 
$\alpha(m_W) = 1/128$,
$\sin^2\theta_{\rm w} = 0.23$,
$m_t = 180 \pm 12 \ \GeVcsq$ \cite{topmass}, 
$m_W = 80.32 \pm 0.19 \ \GeVcsq$ and
$\br(\Ke3) = 4.82 \pm 0.06 \ \%$ \cite{PDG}.
\par
The probability density of $B \equiv \br(\Kpinunu)$,
the branching ratio summed over $\ell = {\rm e},\mu,\tau$,
can be determined using the procedure of 
equation \ref{pdf}

\begin{equation}
\psi(B) 
 = 
  \frac{{\rm d}}{{\rm d} B}
   \int_{\Theta}
    {\rm d}\Theta
     \ P(\Theta)
       \ \lceil \beta(\Theta) < B \rceil
\label{Kth}
\end{equation}
\noindent where $\Theta$ 
is now expanded to include 
$x_t$, $\br(\Ke3)$ and $X^{\ell}_{NL}$ 
as well as the six CKM measurements in Table \ref{table1}.
We assume that $x_t$, $\br(\Ke3)$ and $X^{\ell}_{NL}$ are independent 
and normally distributed about their central values but that the 
values of $X^{\ell}_{NL}$ are completely correlated for the three lepton species.
With these assumptions and the constraints of Table \ref{table1},
there is a 90\% probability that
$0.5 \times 10^{-10} < \br(\Kpinunu) < 2.2 \times 10^{-10}$ and
the most probable value of the branching ratio is 
$ 0.8 \times 10^{-10}$.

\begin{figure}[h]
\begin{center}
%\centerline{\epsfig{file=osc$tmp:5jun1.eps,height=8cm}}
\centerline{\epsfig{file=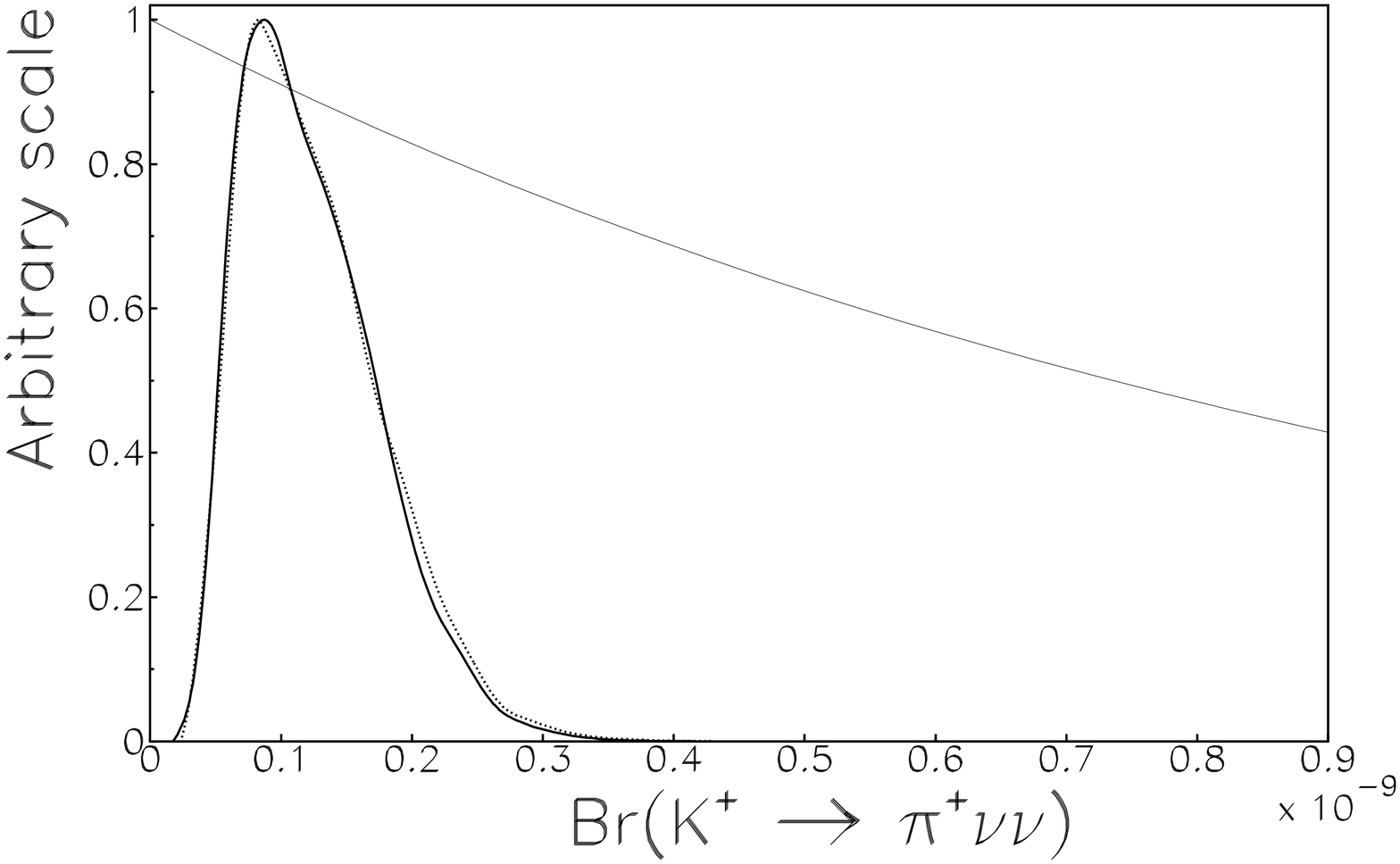,height=6cm}}  % was 8cm
\parbox{15cm}{
\caption{ The dotted curve shows the probability density
         of the \protect$\Kpinunu$
         branching ratio determined from the measurements in Table
         \protect\ref{table1} and equations \ref{Keqn} and \ref{Kth}.
         The thin solid curve is the probability of \protect$\br(\Kpinunu)$
         determined from BNL E787. The thick solid curve is
         the combined probability of \protect$\br(\Kpinunu)$.
         The normalization is arbitrary.
        }
\label{fig3}
}
\end{center}
\end{figure}

The current  experimental upper limit at 90\% confidence level
of $2.4 \times 10^{-9}$ \cite{BNL} from BNL E787 is thus roughly an order of
magnitude higher than the Standard Model prediction. From 
the recommended procedure for the calculation of
confidence levels for Poisson processes with background \cite{PDG},
%we construct 
the probability density for $\br(\Kpinunu)$ 
%from the experimental result
is $  P_e(B) \propto \exp(-BAN_{{\rm K}^+}) $ 
where $A = 0.0027$ is the experimental acceptance and
$N_{{\rm K}^+} = 3.49 \times 10^{11}$ is the number of stopped 
${\rm K}^+$ \cite{BNL}.
The experimental result $P_e(B)$ is compared to the expected range
from equation \ref{Kth} in Figure \ref{fig3}. We also show 
the result of Bayes' theorem for 
the combined
probability for the $\Kpinunu$ branching ratio. As expected the
experimental result has no significant effect on the combined 
probability. The methods used  in the paper should be useful
in the near future as an upgraded version of BNL E787 is expected
to be capable of observing the $\Kpinunu$ decay
assuming the Standard Model is correct \cite{Rob}. In the following section 
we describe the possible effect of such a measurement on the unitarity
triangle.

% --- original location of figure 3

\section{The unitarity triangle}
The unitarity triangle is a convenient way to present the
relations between the CKM matrix elements \cite{PDG,unitri}. The lengths
of the sides of the triangle are given by
$(\cVub\Vud)/(\cVcb\Vcd)$, $(\cVtb\Vtd)/(\cVcb\Vcd)$ and 1.
The side of unit
length conventionally has endpoints at (0,0) and (1,0) in the complex
plane so that the apex of the triangle is given by $(\rho,\eta)$. 

\par
\begin{figure}[h]
\begin{center}
%%%\centerline{\epsfig{file=osc$tmp:6jun1.eps,height=11cm}}
%%\centerline{\epsfig{file=osc$tmp:22july1.eps,height=11cm}}
\centerline{\epsfig{file=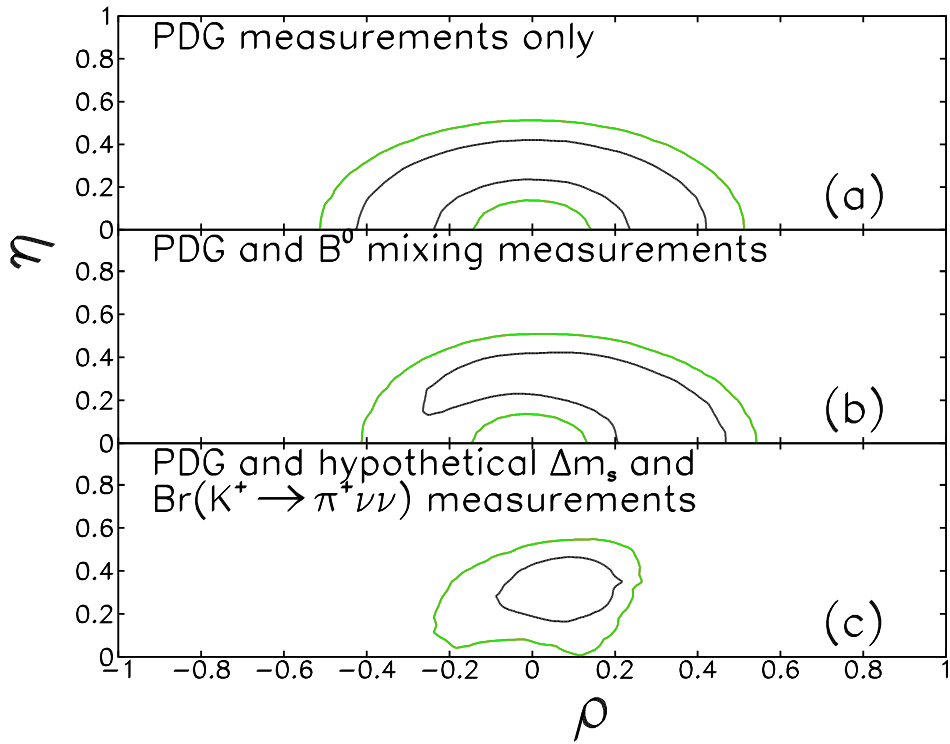,height=11cm}}
%%% CONTOURS#FIG4 makes this figure
\parbox{15cm}{
\caption{
         The 68.3\% (dark) and 95.5\% (light) probability contours 
         for the apex of
         the unitarity triangle determined using 
         (a) the measurements
             listed in Table \protect\ref{table1} only,
         (b) the \protect$\bzbzbar$
             mixing results and Table \protect\ref{table1}, and
         (c) Table \protect\ref{table1} and hypothetical measurements
             of \protect$\dms = 12.5 \pm 0.5\ {\rm ps}^{-1}$ and
             \protect$\br(\Kpinunu) = (1.0 \pm 0.1) \times 10^{-10}$.
         A non-zero $\eta$ would confirm CP violation as
         predicted by the Standard Model.
        }
\label{fig4}
}
\end{center}
\end{figure}

A straightforward generalization of equation \ref{pdf} allows us
to calculate the probability density of $\rho$ and $\eta$,
$\psi(\rho,\eta)$.
Figure \ref{fig4}(a) shows the probability contours of $\psi(\rho,\eta)$
given the measurements in Table \ref{table1}.
The addition
of the $\bzbzbar$ mixing information produces a slight modification
of the contours as shown in Figure \ref{fig4}(b). 
Finally we show the
contours given hypothetical measurements of
$\dms = 12.5 \pm 0.5 \ {\rm ps}^{-1}$ and 
$\br(\Kpinunu) = (1.0 \ \pm \ 0.1) \times 10^{-10}$ in Figure \ref{fig4}(c).
With these two measurements, there would be a 90\% probability that
$0.0077 < |\Vtd| < 0.0106$ and the most probable value of
$|\Vtd|$ would be $0.0088$. As seen in Figure \ref{fig4}(c), 
it would be possible to
confirm the prediction of CP violation ($\eta \not= 0$) by 
the Standard Model at the 95\% confidence level 
with these two particular measurements.

\section{Conclusions}
\par
   Using Bayes' theorem, recent measurements of $\bzbzbar$ mixing
are used to improve upon current estimates of the CKM 
matrix elements with minimal theoretical uncertainty.  
Only $|\Vtd|$ is significantly affected by
these results.  
The impact of current and potential  measurements of 
$\bzbzbar$ mixing and the $\Kpinunu$ branching ratio on the 
individual CKM matrix elements and 
the CKM unitarity triangle is also examined.
\par
This work was supported by US Department of Energy contracts
DE-FG05-92ER40742 and
DE-FC05-85ER250000.

%%%%%%%%%%%%%%%%%%%%%%%%%%%%%%%%%%%%%%%%%%%%%%%%%%%%%%%%%%%%%%%%
%

%
%%%%%%%%%%%%%%%%%%%%%%%%%%%%%%%%%%%%%%%%%%%%%%%%%%%%%%%%%%%%%%%%

\end{document}